\documentclass{cflowproc}

\usepackage{natbib}

\begin{document}


\shortauthors{Peterson et al.}     
\shorttitle{Constraints on Cooling Flows} 

\title{High Resolution X-ray Spectroscopic Constraints on Cooling-Flow Models}   

\author{John R. Peterson,\affilmark{1} Steven M. Kahn,\affilmark{1} Frits B. S. Paerels,\affilmark{1}   
Jelle S. Kaastra,\affilmark{2} Takayuki Tamura,\affilmark{2}
Johan A. M. Bleeker,\affilmark{2} Carlo Ferrigno,\affilmark{2}                
and J. Garrett Jernigan\affilmark{3}}

\affil{1}{Columbia Astrophysics Laboratory, Columbia University, 550 W 120th St., NY, NY 10027, USA}   
\and
\affil{2}{SRON National Institute for Space Research, Sorbonnelaan 2, 3584 CA
				Utrecht, The Netherlands}
\and                                
\affil{3}{Space Sciences Laboratory, University of California, Berkeley, CA
				94720, USA}


\begin{abstract}
We present XMM-Newton Reflection Grating Spectrometer observations of X-ray
clusters and groups of galaxies. We demonstrate the failure of the standard
cooling-flow model to describe the soft X-ray spectrum of clusters of
galaxies. We also emphasize several new developments in the study of the soft
X-ray spectrum of cooling flows.  Although there is some uncertainty in the
expected mass deposition rate for any individual cluster, we show that high
resolution RGS spectra robustly demonstrate that the expected line emission
from the isobaric cooling-flow model is absent below 1/3 of the background
temperature rather than below a fixed temperature in all
clusters. Furthermore, we demonstrate that the best-resolved cluster spectra
are inconsistent with the predicted shape of the differential luminosity
distribution and the measured distribution is tilted to higher
temperatures. These observations create several fine-tuning challenges for
current theoretical explanations for the soft X-ray cooling-flow problem.
Comparisons between these observations and other X-ray measurements are discussed. 

\end{abstract}





\section{The Cooling-Flow Model}

A long-standing theoretical prediction is that the intracluster medium in the cores of
galaxy clusters should cool by emitting X-rays in less than a Hubble time
(\citealt{fabian}; \citealt{cowie}; \citealt{mathews}).   If the
cooling proceeds inhomogenously (\citealt{nulsen1}; \citealt{johnstone2}), a range of temperatures is likely to exist in
the cooling plasma.  Straightforward thermodynamic arguments show that at
constant pressure the particular distribution of plasma temperatures is
described by the differential luminosity distribution, 

\begin{equation}
\frac{dL}{dT} = \frac{5}{2} k \frac{\dot{M}}{\mu m_p}
\end{equation}

\noindent
where $k$ is Boltzmann's constant, $\mu m_p$ is the mean mass per particle,
and $\dot{M}$ is the mass deposition rate.  This yields a {\it unique} X-ray
spectrum for an assumed set of elemental abundances and a given maximum temperature
where the cooling is assumed to begin.  Inhomogenous cooling of this sort is
broadly called a {\it cooling flow}.

We can apply Equation 1 to the X-ray spectrum by using standard plasma codes
to predict the energy-dependent luminosity at a given temperature.  There
are three major assumptions:  First, it is normally assumed that the total luminosity emitted from the high
temperature plasma is not substantially different than that observed in
X-rays.  If this were not true or if the plasma was indirectly transfering its
thermal energy to another source, then this would invalidate the model.
Second, we assume that there is no substantial heating that
would modify the thermodynamic history of the plasma.  Third, we assume
that the bulk of the cooling volume is approximately in a steady-state, so
that the cooling flow is not dramatically changing during a characteristic
cooling time.  Although many modifications and theoretical challenges to this
simple model have been discussed for more than 25 years, it was not known
until recently if the X-ray spectrum deviated significantly from this simple cooling-flow model.  

In order to measure a differential luminosity distribution at keV temperatures
like that in Equation 1, emission lines from individual ions need to be
measured.  In particular, it is important
to measure emission lines from the Fe L series, which are ions
that happen to have their ionization potentials in the temperature range near
and below cluster virial temperatures.  This is shown in Figure 1 where the
relative charge state abundance (elemental abundance times fractional ionic
abundance) of various ions is plotted.  However, measuring the relative
strength of emission from individual Fe L ions is only possible to do in detail
with resolving powers ($\frac{\lambda}{\Delta \lambda}$) above 100.  These
resolving powers are not achievable with non-dispersive CCD experiments like
ASCA/SIS, XMM-Newton/EPIC, and Chandra/ACIS, although some information can be
obtained from the shape of the unresolved Fe L emission line complex.  The
Reflection Grating Spectrometer (RGS) on XMM-Newton, however, is capable of high-resolution spectroscopy for
sources with spatial extent similar to many cooling flows.

\begin{figure}
\plotone{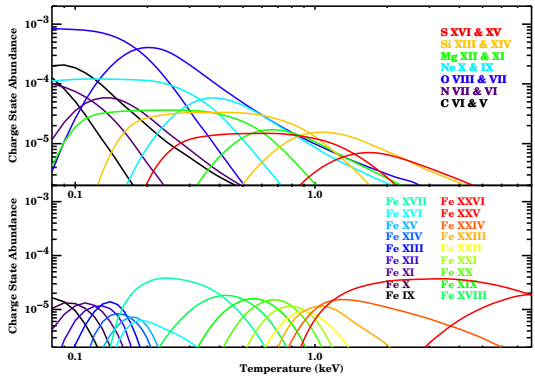}
\figcaption{Fractional abundance of a given ion
  plotted against temperature in keV (taken from \citealt{arnaud}).  The
  fractional abundance is multiplied by the abundance of that element relative
  to hydrogen in the solar neighborhood (\citealt{anders}).  The top plot shows the helium and
  hydrogen-like charge states for several elements.  The bottom plot shows a number of
  charge states of iron.\label{peterson:f1}}
\end{figure}

The RGS is a dispersive spectrometer that contains a grating array that is placed
behind the XMM mirrors.  The grating array deflects photons to a CCD bench.   The RGS was designed to have a
relatively high dispersion of 3 degrees for the soft X-ray band in order to
compensate for the fact that the XMM mirrors would blur sources by about 10
arcseconds (\citealt{kahn}).  Cooling flows happen to have a similar size to the
XMM blurring, so that observations of cooling flows also benefit
from the large dispersion angles.  The dispersion of an incident X-ray is
determined by both the wavelength of the photon and the angle it hits the
grating array.  Therefore, for a source with significant spatial extent like a
cluster, the coupling of the sky position with the wavelength has to be
modeled.  A Monte Carlo approach has therefore been used to analyze extended
source RGS observations.  Some discussion of this can be found in \cite{peterson2}.

\section{Spectroscopic Constraints on Cooling Flows}

The initial application of the standard cooling-flow model to a high resolution
spectrum of a cluster of galaxies is shown in Figure 2.  The top panel
shows the spectral prediction for a model based on Equation 1.  The second
panel shows the comparison between the model (blue) and the spectrum of Abell 1835 (red).
Notably absent in the data are Fe XVII lines.  Finally, the bottom panel shows a model where the
emission below 3 keV is suppressed.  This fairly simple model of the X-ray
emission seems to describe the data quite well.  The plasma appears to match
the cooling flow model betwen 3 keV and the maximum cluster temperature of 8
keV but not below 3 keV.  The obvious theoretical question is why it appears close to
matching the cooling flow model at high temperatures, but not at low
temperatures.  It is clearly not an isothermal spectrum, but it is also
inconsistent with a complete cooling flow model.  Full details of this
comparison can be found in \cite{peterson1}.  Other grating observations
contain similar results (\citealt{tamura}; \citealt{kaastra}; \citealt{tamura2}; \citealt{xu}; \citealt{sakelliou}).

\begin{figure}
\plotone{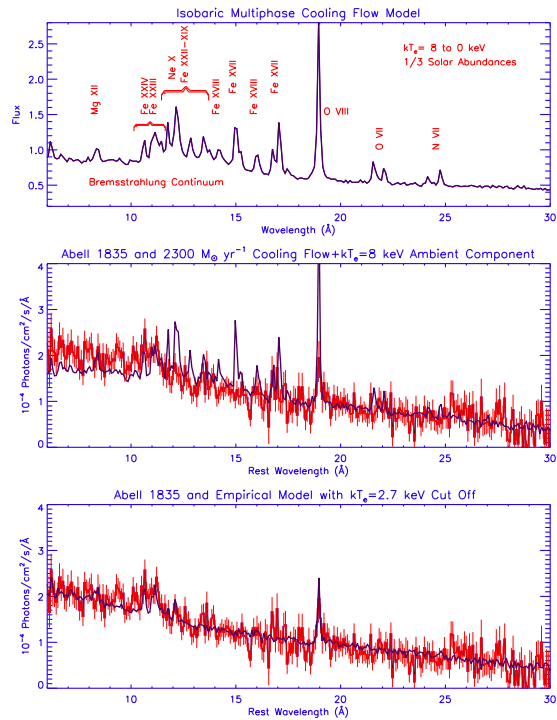}
\figcaption{Comparison of the cooling flow model to Abell 1835.  The top panel shows the cooling flow model based on Equation 1 in the text.  Note the numerous Fe L lines predicted in the 10 to 17 $\mbox{\AA}$ region.  The middle panel shows the comparison of that model with the data from the galaxy cluster Abell 1835 which was thought to host a
  massive cooling flow.  The bottom panel shows the cooling flow model with the emission suppressed below 3 keV.\label{peterson:f2}}
\end{figure}

\begin{figure}
\plotone{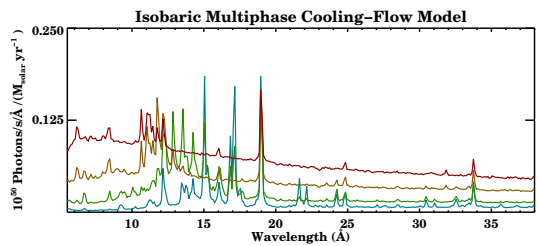}
\figcaption{The emission spectrum of the standard cooling flow
  model broken up into several temperature bins.  The red line shows the 3 to
  6 keV spectrum, the yellow shows the 1.5 to 3 keV spectrum, the green shows
  the 0.75 to 1.5 keV spectrum and the blue shows the 0.375 to 0.75 keV
  spectrum.  The normalization of each of the components is adjusted to see
  where the cooling flow model fails.\label{peterson:f3}}
\end{figure}

If we decompose the spectrum in the top panel in Figure 2 into different
temperature ranges, we arrive at the spectra in Figure 3.  Here we plot the
6 to 3 keV (red), 3 to 1.5 keV (yellow), 1.5 to 0.75 keV (green), and 0.75 to
0.375 keV spectrum (blue).  Then if we fit the observed X-ray spectra while
allowing the relative normalization of each of these four spectra to vary, we
can fit the X-ray spectra of 14 cooling-flow clusters of galaxies quite well.
This is shown in the spectral fits in Figures 4-6.  These fits also use a
spatial model for the cluster emission folded through the Monte Carlo for
the RGS.  The red line shows the fit, the blue histogram is the data, and the
green line is the standard cooling flow model with no adjustments.  The
cooling flow model deviates significantly from the data in a similar manner to
the Abell 1835 comparison in Figure 2.  The low temperature emission is not
present or significantly reduced in the cluster spectra.  Full details of the
model and analysis method can be found in \cite{peterson3}.

Figure 7 shows the
differential luminosity of each of the four spectra (proportional to the total
normalization given by the fit) against the temperature.  Each cluster is
shown in a different color and the actual detections are connected by a line.
If the cooling flow model were correct, then the points should line close to
the line $y=1$.  Figure 8 shows the same plot but it is plotted against the
temperature relative to the maximum temperature.  The points appear roughly
consistent with a power law in fractional temperature where the index is near
1 to 2.

\begin{figure}
\plotone{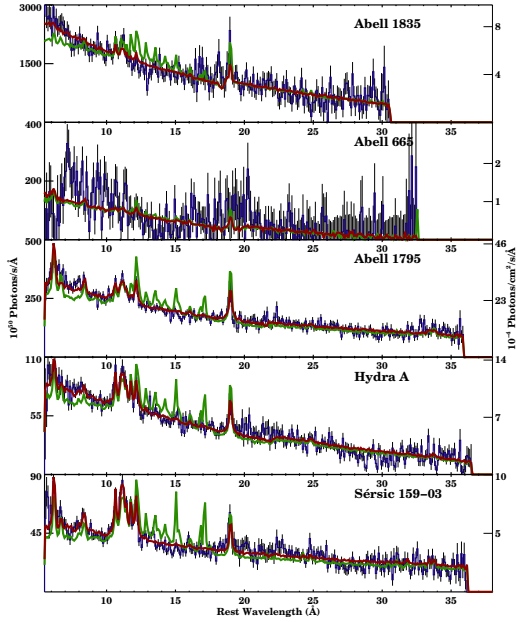}
\figcaption{Fluxed spectra of several hot (about 5 keV) clusters of galaxies.
  The red line is an empirical fit allow the normalization of the cooling flow
  model for several temperature ranges to be adjusted.  The green line is the
  standard cooling flow model with no adjustments.  The model fails
  most strongly from Fe XVII lines, which implies that the lowest temperature
  emission is absent.\label{peterson:f4}}
\end{figure}

\begin{figure}
\plotone{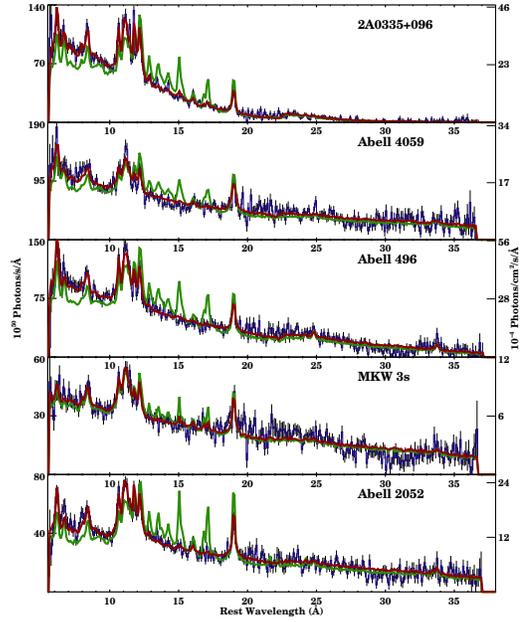}
\figcaption{The same plot as Figure 4 but with medium
  temperature (3 to 5 keV) clusters.\label{peterson:f5}}
\end{figure}

\begin{figure}
\plotone{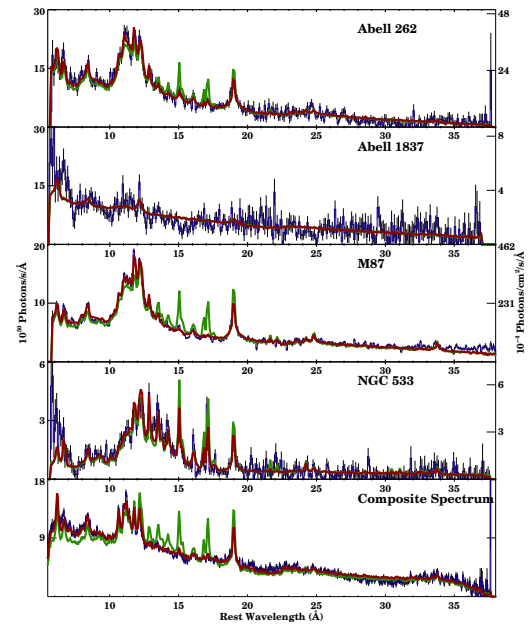}
\figcaption[Cooling clusters]{The same plot as Figure 4 and 5 but with cool
  (less than 3 keV)
  clusters and groups.\label{peterson:f6}}
\end{figure}

Figure 7 and 8 demonstrate several aspects of the failure of the cooling-flow
model.  First, it is clear that significant quantities of plasma below the
maximum temperature always exists.
Second, the model appears to fail at a fraction of the maximum temperature
rather than a fixed value as shown by the luminosity distribution crossing the
line $y=1$ near $T=\frac{1}{3}$ in Figure 8.  Furthermore, the model fails
in its shape (it is not horizontal) since it is tilted towards higher temperatures and the
normalization at low temperatures is not the only discrepancy.  The overall normalization, however, is somewhat difficult to
interpret in the absence of a new theoretical model, since a new model would
probably add or subtract heat to the system which affects the normalization.  There is also significant
scatter in both plots.  It is not clear whether this is due to subtle fitting
degeneracies or is a real difference between cooling flows.  We also cannot be
certain if the empirical model, the power law in fractional temperature,
continues to arbitrarily low temperatures.  It is clear, however, that the
model is off by as much as an order of magnitude at the lowest temperatures.
Finally, it appears that the simple temperature cut-off model used in Figure 2
might be oversimplified.  The best-resolved spectra like that for 2A0335+096
has emission lines from Fe XXIV through XVIII.  This has forced us to use a more complex model
than the cut-off model to match the lowest temperature emission.   Similarly,
just reducing the overall normalization and therefore the implied mass
deposition, $\dot{M}$, also does not describe these observations.  Clearly, the many aspects of these observations should
offer some clue to the solution to this problem.

\begin{figure}
\plotone{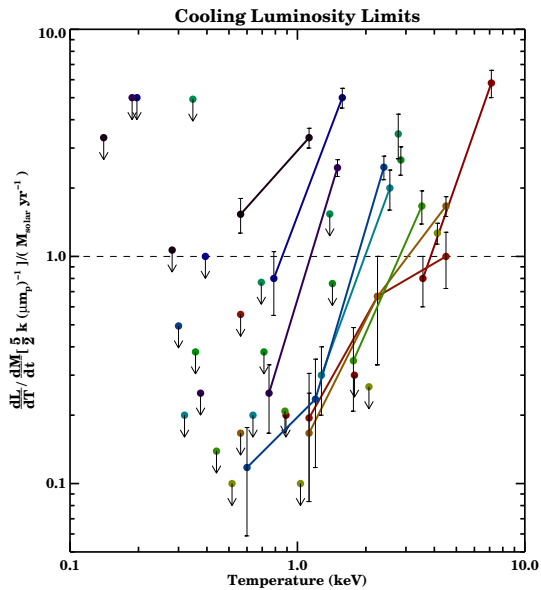}
\figcaption{A comparison of the luminosity in specific temperature ranges
  to the cooling flow model.  See the text for details.\label{peterson:f7}}
\end{figure}

\begin{figure}
\plotone{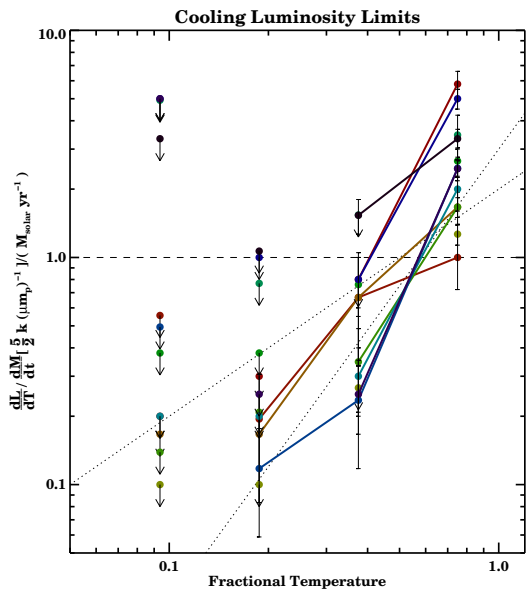}
\figcaption{Same as Figure 7 but plotted as a function of the fraction of the
  maximum temperature.\label{peterson:f8}}
\end{figure}

\section{Theoretical Challenges}

There are many ideas for possible mechanisms which would either add or
subtract heat from cooling flows (see e.g., \citealt{peterson1};
\citealt{fabian3}; \citealt{boehringer2}; \citealt{sasaki};
\citealt{makishima}; \citealt{peterson3}; \citealt{mathews2}). Promising scenarios are documented in these
proceedings and include dust cooling (\citealt{fabian3}; \citealt{mathews2}), cold cloud interface
mixing (\citealt{begelman}; \citealt{fabian3}),
interactions with cosmic rays (\citealt{gitti}), heating through processes associated with
relativistic AGN outflows (\citealt{rosner}; \citealt{binney};
\citealt{churazov}; \citealt{brueggen}; \citealt{quilis}; \citealt{david};
\citealt{nulsen3}; \citealt{kaiser};  \citealt{ruszkowski}; \citealt{soker}; \citealt{brighenti}), cluster merging (\citealt{markevitch2}), or significant heat transfer
between the outer region and cores of clusters (\citealt{tucker};
  \citealt{stewart}; \citealt{zakamska}; \citealt{voigt};  \citealt{fabian5};
  \citealt{soker2}).  Less interesting explanations that might involve subtle errors in the application of Equation
1, such as a gross error in the plasma codes, are probably not relevant.  This
is likely not to be the case since some cataclysmic variables are observed to
have X-ray spectra almost exactly like that in the top panel in Figure 2
(\citealt{mukai}).  This is understood by the idea that the accretion of
material from the secondary to the white dwarf where it must cool down to the
photospheric temperature essentially proceeds the same way as a classic cooling flow.

Generally, there are three major requirements for any successful
model.  First, the average heating or additional cooling power should be
comparable to the X-ray luminosity from cooling flows.  Most calculations have
shown that this could be true for a number of different models.  The second and
somewhat more confusing requirement is that the mechanism should work more
effectively at lower temperatures or should begin to operate right before
complete cooling occurs.  The isobaric cooling time is proportional
approximately to the second power of temperature.  So if the cooling starts to
be disrupted characteristically at $\frac{1}{3}$ of the maximum temperature,
why does it stop when it is $\frac{8}{9}$ of the way to completion?
Finally, a successful model must also add or subtract the heat at the
right time at all spatial positions.  The observed spatial distribution of the X-ray emission, however, is somewhat complicated
as we discuss in the following section.  The global problem, however, is quite
clear-- the X-rays from the lowest temperatures are not present in the entire
cooling flow volume.

\section{Comparison with other X-ray observations}

Some discussion on the nature of fitting spatially-resolved collisional X-ray
spectra is warranted, since the comparison of these observations with those
derived from other X-ray instruments can be complicated and was under frequent
discussion at this conference.  If the ICM is optically-thin to its own
radiation and is in collisional equilibrium then its differential luminosity
distribution is described by two functions,

\begin{equation}
\frac{d^2 L}{d T d \Omega} = f_1 (T,\Omega)
\end{equation}

\begin{equation}
\frac{d A_i}{d \Omega} = f_2 (\Omega)
\end{equation}

\noindent
where the first function $f_1$ describes the differential luminosity
distribution as a function of temperature, $T$, at each solid angle element
$d\Omega$ and the second function $f_2$ specifies an abundance map for each
element.  The measurement of other quantities, such as a density or volume, are always
indirectly derived from these two functions.  Note also that the function, $\frac{d^2 A_i}{d T d \Omega}$, is not actually observable in
practice.  This is because a plasma consisting of multiple temperatures will
emit line emission that can be separately measured from each temperature
region but the continuum emission from each temperature region will just
contribute to one overall continuum.  Even if the plasma were divided into
different temperatures each with their own metallicity, only one overall
metallicity can be measured for any one spectrum.

The ability to measure a temperature distribution, $\frac{dL}{dT}$, is
determined by the number of counts in the spectrum, the resolution of the
instrument, and fundamental atomic physics issues.  For a high resolution
spectrum with a large number of counts, the X-ray temperature distribution
probably cannot be divided into much more than 8 or so temperature bins.  This
is because there are only so many strong emitting X-ray ions.  The Fe K and L
series, for example, have 10, but their overlap in specific temperatures ranges
makes disentangling the temperature distribution in great detail difficult.
Line ratios between emission lines from the same ion are only weakly temperature
sensitive for the normal temperature ranges that the ion is produced.

With a typical RGS cluster observation of about 30,000 source photons, we have
typically found that deviding the X-ray temperature range into 4 logarithmic
bins usually provides a robust solution.  With more photons, we expect that a
slightly more detailed analyses can be performed.  With the non-dispersive CCD
instruments of EPIC and Chandra less information can be extracted.  Individual
X-ray ions are not detected precisely, but their emission is blended
significantly.  It appears that usually determining much more than two
temperatures (or two temperature ranges) is extremely difficult without a very large number of counts.

The RGS observations place a different set of constraints on the spatial
distribution of the luminosity distribution, $\frac{dL}{d \Omega}$ than EPIC
or Chandra observations.  The ability to do this, of course, depends on the spatial
resolution of the instrument provided enough X-ray photons are collected.
Therefore, Chandra can constrain the angular luminosity distribution at the
arcsecond scale, whereas EPIC can constrain the distribution at the several
arcsecond scale.  RGS observations can utilize the cross-dispersion
information (the distance photons are detected perpendicular to the dispersion
axis of the spectrometer) to measure the spatial distribution almost as good
as the EPIC resolution.  The shapes of line profiles at least in principle
can constrain the distribution in the dispersion direction.

The ability to measure the total two-dimensional function, $\frac{d^2 L}{d T d
  \Omega}$ is obviously complicated and probably is different for every
  observation of every cluster.  
  We may have to guess at what the true distribution or what the shape of
this function is for cooling flows.  Making conclusive statements about isothermality or a preferred
 formulation of the temperature distribution is very challenging and depends
 heavily on the capabilities of that instrument.
 All of the instruments, however, have generally shown an average temperature decline.
 The RGS observations presented here show evidence for a steep differential luminosity
 distribution of order $T^1$ or $T^2$ if averaged over the entire cooling flow.
 The deepest EPIC observations show steep multiphase distributions (possibly
 power laws with indices
 greater than $T^2$) with overall temperature contrasts that are probably
 greater than a factor of two (\citealt{kaastra2}; \citealt{lewis}).
 Chandra thermal images have shown non-spherical temperature variations
 at levels less than 50 $\%$ (\citealt{sanders}).  We might speculate that all the observations
 have given us a picture that the combination of an overall temperature decline
 and a steep multiphase distribution ($T^3$) at each radius combines to give an
 overall differential distribution near to what has been measured with the
 RGS ($T^1$ or $T^2$).  This, however, has not been fully explored nor is it clear
 how much difference there is between any of the clusters.

 The abundance map, $\frac{dA_i}{d \Omega}$, is only beginning to be measured since it depends
 heavily on getting the temperature distribution correct.  Furthermore,
 deviations from collisional equilbrium have yet to be fully explored.  In
 addition, the optically-thin assumption as well could be violated by resonant
 scattering but the effect is either observationally controversial (see e.g.,
 \citealt{boehringer2}; \citealt{xu}; \citealt{churazov2}) or could be
 different for every cluster due to varying velocity fields.

\section{Future Work}
As we have argued in the previous section, there is much more work to be done
in clarifying the observational situation in X-rays.  Similarly, the
connection of these observations to other wavelengths is very incomplete.  In
particular, maps that correlate the quantity of missing X-rays with the
detections of cold material will be important (\citealt{edge};
\citealt{edge2}).  Theoretical studies should continue to build towards
a final solution.  

It has been previously discussed that cooling flows of some
form are necessary to cool galaxies to form stars before they get incorporated
into much larger structures (\citealt{white}).  Unless the solution to the
soft X-ray cooling flow problem is endemic to cluster scales, it could have far-reaching
consequences for structure formation.

\acknowledgements
JRP would like to thank the conference organizers and the attendees for a very
successful conference.  Work on the RGS at Columbia University and U. C. Berkeley is supported by NASA.  The laboratory for Space Research,
Utrecht is supported by NWO, the Netherlands Organization for Scientific
Research.

\begin{thebibliography}{}

\bibitem[Anders \& Grevesse(1989)]{anders} Anders, E. \& Grevesse, N. 1989, Geochimica et Cosmochimica Acta 53, 197.

\bibitem[Arnaud \& Raymond(1992)]{arnaud}Arnaud, M. and J. Raymond, 1992, ApJ 398, 394.

\bibitem[Begelman \& Fabian(1990)]{begelman} Begelman, 
M. \& Fabian, A. C. 1990, MNRAS 244, 26.

\bibitem[Binney \& Tabor(1995)]{binney} Binney, J. \& Tabor, G. 1995, MNRAS
  276, 663.

\bibitem[B\"ohringer et al.(2002)]{boehringer2} 
B\"ohringer, H., Matsushita, K., Churazov, E., Ikebe, Y., \& Chen, Y. 2002, A\&A 382, 804.

\bibitem[Brighenti \& Mathews(2003)]{brighenti} Brighenti, F. \& Mathews,
  W. G. 2003, ApJ 587, 580.

\bibitem[Br\"uggen \& Kaiser(2001)]{brueggen} Br\"uggen, M. \&  Kaiser, C. R. 2001, MNRAS 325, 676.

\bibitem[Churazov et al.(2001)]{churazov} Churazov, E., Sunyaev, R., Forman,
  W., B\"ohringer, H. 2001, ApJ 554, 261.

\bibitem[Churazov et al.(2003)]{churazov2} Churazov, E., Forman, W., Jones,
  C., Sunyaev, R., \& B\"ohringer, H. 2003, MNRAS submitted.

\bibitem[Cowie \& Binney(1977)]{cowie} Cowie, L. L. \& 
 Binney, J. 1977, ApJ 215, 723.

\bibitem[David et al.(2001)]{david} David, L. P., Nulsen, P. E. J., McNamara, B. R., Forman, W., Jones, C., Ponman, T., Robertson, B., \& Wise, M. 2001, ApJ 557, 546.

\bibitem[Edge \& Frayer(2003)]{edge2} Edge, A. C. \& Frayer, D. T. 2003, ApJ
  594, 13.

\bibitem[Edge et al.(2002)]{edge} Edge, A. C. et al. 2002, MNRAS 337, 49.

\bibitem[Fabian, Voigt, \& Morris(2002)]{fabian5} Fabian, A. C., Voigt, L. M.,
  \& Morris, R. G. 2002, MNRAS 335, L71.

\bibitem[Fabian et al.(2001)]{fabian3} Fabian, A. C., Mushotzky, R. F.,
  Nulsen, P. E. J., \& Peterson, J. R. 2001, MNRAS 320, 20.

\bibitem[Fabian \& Nulsen(1977)]{fabian} Fabian, A. 
C. \& Nulsen, P. E. J. 1977, MNRAS 180, 479.

\bibitem[Gitti, Brunetti, \& Setti(2002)]{gitti} Gitti, M., Brunetti, G., \&
  Setti, G., A\&A 386, 456.

\bibitem[Johnstone et al.(1992)]{johnstone2} Johnstone, R. M., Fabian, A. C., Edge, A. C., \& Thomas, P. A. 1992, MNRAS 255, 431-440.

\bibitem[Kaastra et al.(2003)]{kaastra2} Kaastra, J. S. et al., this
  conference.

\bibitem[Kaastra et al.(2001)]{kaastra} Kaastra, J. S., Ferrigno, C., Tamura, T., Paerels, F. B. S., Peterson, J. R., \& Mittaz, J. P. D. 2001, A \& A 365, 99.

\bibitem[Kaiser \& Binney(2002)]{kaiser} Kaiser, C. \& Binney, J. 2002, MNRAS
  338, 837.

\bibitem[Kahn \& Hettrick(1985)]{kahn} Kahn, S. M. \& Hettrick, M. C. 1985, in
  ESA Proceedings of a ESA Workshop for a Cosmic X-ray Spectroscopy Mission,
  Paris: ESA, 237. 

\bibitem[Lewis et al.(2003)]{lewis} Lewis, A. D., Buote, D. A., Mathews,
  W. G., \& Brighenti, F. 2003, this conference.

\bibitem[Makishima et al.(2001)]{makishima} Makishima, K. et al. 2001, PASJ
  53, 401.

\bibitem[Markevitch et al.(2001)]{markevitch2} Markevitch, M., Vikhlinin, A.,
  \& Mazzotta, P., 2001, ApJ 562, L153.

\bibitem[Mathews \& Bregman(1978)]{mathews} Mathews, W. G. \& Bregman,
  J. N. 1978, ApJ 224, 308.

\bibitem[Mathews \& Brighenti(2003)]{mathews2} Mathews, W. G. \& Brighenti,
  F. 2003, ApJ 590, 5.

\bibitem[Mukai et al.(2003)]{mukai} Mukai, K., Kinkhabwala, A., Peterson,
  J. R., Kahn, S. M, \& Paerels, F. B. S. 2003, ApJL 586, 77.

\bibitem[Nulsen(2002)]{nulsen3} Nulsen, P. E. J., David, L. P., McNamara, B. R., Jones, C., Forman, W. R., Wise, M. 2002, ApJ 568, 163.

\bibitem[Nulsen(1986)]{nulsen1} Nulsen, P. E. J. 1986, MNRAS 221, 377.

\bibitem[Peterson et al.(2003b)]{peterson3} Peterson, J. R. et al. 2003a, ApJ
  590, 207.

\bibitem[Peterson, Jernigan \& Kahn(2003a)]{peterson2}  Peterson, J. R.,
  Jernigan, J. G. \& Kahn, S. M. 2003b, Proc. SPIE 4847.

\bibitem[Peterson et al.(2001)]{peterson1} Peterson, J. R. et al. 2001a, A \& A 365, 104.

\bibitem[Quilis(2001)]{quilis} Quilis, V., Bower, R. G., \& Balogh, M. L. 2001, MNRAS 328, 1091.

\bibitem[Rosner \& Tucker(1989)]{rosner} Rosner, R. \& Tucker, W. 1989, ApJ 338, 761.

\bibitem[Ruszkowski \& Begelman(2002)]{ruszkowski} Ruszkowski, M. \& Begelman,
  M. C. 2002, ApJ 586, 384.

\bibitem[Sakelliou et al.(2002)]{sakelliou} Sakelliou, I. et al., A \& A 391,
  2002.

\bibitem[Sanders \& Fabian(2003)]{sanders} Sanders, J. S. \& Fabian,
  A. C. 2003, this conference.

\bibitem[Sasaki \& Yamasaki(2002)]{sasaki} Sasaki, S. \& Yamasaki, Y. 2002, PASJ 54, 1.

\bibitem[Soker(2003)]{soker2} Soker, N. 2003, MNRAS 342, 463.

\bibitem[Soker \& David(2003)]{soker} Soker, N. \& David, L. P. 2003, ApJ 589, 77.

\bibitem[Stewart et al.(1984)]{stewart} Stewart, G. C., Fabian, A. C., Nulsen, P. E. J., \& Canizares, C. R. 1984, ApJ 278, 536.
\bibitem[Tamura et al.(2001a)]{tamura} Tamura, T. et al. 
2001, A \& A 365, 87.

\bibitem[Tamura et al.(2001b)]{tamura2} Tamura, T., Bleeker, J. A. M., Kaastra, J. S., Ferrigno, C., \& Molendi, S. 2001, A \& A 379, 101.

\bibitem[Tucker \& Rosner(1983)]{tucker} Tucker, W. H. \& Rosner, R. 1983, ApJ 267, 547.

\bibitem[Voigt et al.(2002)]{voigt} Voigt, L. M., Schmidt, R. W., Fabian, A. C., Allen, S. W. \& Johnstone, R. M. 2002, MNRAS 355, 7.

\bibitem[White \& Rees(1978)]{white} White, S. D. M. \& Rees, M. J. 1978,
  MNRAS 189, 341.

\bibitem[Xu et al.(2002)]{xu} Xu, H. et al. 2002, ApJ 576, 600.

\bibitem[Zakamska \& Narayan(2001)]{zakamska} Zakamska, N. \& Narayan,
  R. 2002, ApJ 582, 162.
























\end{thebibliography}

\end{document}